\documentclass{osa-article}

\journal{oe}

\articletype{Research Article}
\usepackage{siunitx}
\usepackage{textcomp}
\usepackage{multirow}

\newcommand{\figref}[2]{Fig.~\ref{#1}~\textbf{#2}}
\newcommand{\equaref}[1]{Eq.~\ref{#1}}

\begin{document}

\title{Designing of strongly confined short-wave Brillouin phonons in silicon waveguide periodic lattices}

\author{Roberto O. Zurita$^{*}$, Gustavo S. Wiederhecker, Thiago P. Mayer Alegre$^{\dagger}$}
\address{Applied Physics Department and Photonics Research Center, ``Gleb Wataghin'' Institute of Physics, University of Campinas, 13083-859, Campinas, SP, Brazil}
\email{*zurita@ifi.unicamp.br, $\dagger$ alegre@unicamp.br}


\begin{abstract*}
We propose a feasible waveguide design optimized for harnessing Stimulated Brillouin Scattering with long-lived phonons. The design consists of a fully suspended ridge waveguide surrounded by a 1D phononic crystal that mitigates losses to the substrate while providing the needed homogeneity for the build-up of the optomechanical interaction. The coupling factor of these structures was calculated to be $\SI{0.54}{(W.m)^{-1}}$ for intramodal backward Brillouin scattering with its fundamental TE-like mode and $\SI{4.5}{(W.m)^{-1}}$ for intramodal forward Brillouin scattering. The addition of the phononic crystal provides a \SI{30}{\dB} attenuation of the mechanical displacement after only five unitary cells, possibly leading to a regime where the acoustic losses are only limited by fabrication. As a result, the total Brillouin gain, which is proportional to the product of the coupling and acoustic quality factors, is nominally equal to the idealized fully suspended waveguide.
\end{abstract*}

\section{Introduction}
Stimulated Brillouin Scattering (SBS) is a non-linear phenomenon that arises from the coherent coupling between light and sound. Thermal noise driven acoustic waves scatters a strong  pump wave to a red-shifted Stokes wave, mechanical vibrations in the medium are then further stimulated by the optical forces resulting from the beating between  pump and scattered waves, creating thus a feedback-loop  that can efficiently transfer energy from the pump to the Stokes wave~\cite{Boyd2003}. Experimental studies of SBS began with optical fibers due to their long interaction length and relatively small optical mode areas~\cite{Ippen1972,Hill1976,Rowell1979}. With the advances in microfabrication, it was possible to bring SBS to the subwavelength-confinement scale, where enhanced effects of radiation pressure granted a higher power transfer between pump and probe, despite the small interaction lengths~\cite{Rakich2012,Wiederhecker2019,Safavi-Naeini2019,Eggleton2019}. Since its first experimental demonstration~\cite{Chiao1964}, not long after the first  laser demonstration, SBS-based devices have been explored in many applications ranging from sensors~\cite{xu2016full,Zarifi2018}, notch filters~\cite{Byrnes2012,Feng2018,Kittlaus2018}, RF-signal processing~\cite{Choudhary2017,shin2015control}, enhanced gyroscope~\cite{Li2017}, lasers~\cite{Grudinin2009, Tomes2009,Li2012,Morrison2017, Otterstrom2018,Gundavarapu2019}, and as a platform for exquisite studies~\cite{Pant2012, Lai2019,Wang2020}.

However, mechanical confinement still remain a challenge for achieving large Brillouin gain in CMOS compatible integrated waveguides. Due to the lower acoustic speed of silica when compared to material such as silicon or silicon nitride, the vibrations excited in the device layer is partially transmitted to the substrate at their interface~\cite{Gyger2020,van2015interaction}, as opposed to more exotic platforms like soft glasses in which acoustic waves are trapped by total internal reflection~\cite{kabakova2013narrow}.
For monolithic platforms one popular solution to this problem is to partially~\cite{van2015interaction} or completely undercut the substrate~\cite{van2015net,kittlaus2016large} to suspend the waveguides and isolate them from the substrate . While suspension is very effective at suppressing mechanical losses, it either requires a critical control of the residual anchoring \cite{van2015interaction} -- which might affect distinct mechanical modes in different proportions -- or rely on periodic anchoring points that ease the fabrication process~\cite{kittlaus2016large,kittlaus2017chip,van2015net}.In this latter approach, the structural integrity of the suspended section can easily come into conflict with the necessary optomechanical interaction build-up length. When the characteristic gain length is comparable to the acoustic decay length the optomechanical interaction needs to build-up several acoustic decay lengths before reaching its expected performance~\cite{Wolff2015}. This, however, is particularly difficult for Backwards Brillouin Scattering where the larger group velocity leads to a larger acoustic decay length. Although this regime has not yet been reached in silicon devices, it could become a reality with the increase in both coupling factor ($G_B/Q_m$) and mechanical quality factor ($Q_m$). Reported $Q_m$ in silicon waveguides~\cite{van2015interaction,van2015net,kittlaus2016large} are still bellow silicon's material limit~\cite{Renninger2018} and any scattering and two-level systems induced by microfabrication process as shown by low-temperature experiments~\cite{MacCabe2019,Ren2020}.

One effective approach to overcome these limitations imposed by anchoring points is to explore a phononic crystal cladding along the suspended membrane. The crystal cladding prevents the mechanical vibration from leaving the active region and leaking to the substrate. It also shields the mechanical mode from any perturbation from the necessary periodic anchoring, reconciling the structural integrity with homogeneity needed for long-range Brillouin interactions. Indeed, previous works effectively explored this principle in the particular case of forward Brillouin scattering, where the mechanical waves have a null acoustic wavevector, $\beta_m\approx0$ ~\cite{santos2017hybrid,shin2015control,Dehghannasiri2017}. However, silicon-based waveguide structures favoring the regime of backward Brillouin scattering, which has been shown to enable important functionalities~\cite{Zarifi2018, Feng2018,Choudhary2017,Li2017,Pant2012} and, ultra-low mechanical dissipation~\cite{Renninger2018} remains largely unexplored.

In this work, we investigate backward Brillouin interaction in a CMOS compatible waveguide design comprised of a silicon rectangular waveguide on top of a suspended silicon membrane. We overcome the impact of using anchoring points by surrounding the Brillouin-active waveguide core with a 1-D phononic crystal cladding that supports a full phononic band-gap -- both shear and longitudinal polarizations -- for short-wavelength acoustic modes. We start by highlighting the differences between the mechanical modes behavior in Forward (FBS) and Backwards (BBS) Brillouin scattering. The design process is then deeply discussed showing the phononic crystal cladding optimization for FBS and BBS while discussing the advantages of such design strategy and its expected performance compared to other waveguide design.

\section{Backward and Forward Scattering}

\begin{figure}[ht!]
    \centering
    \includegraphics[width=1\columnwidth]{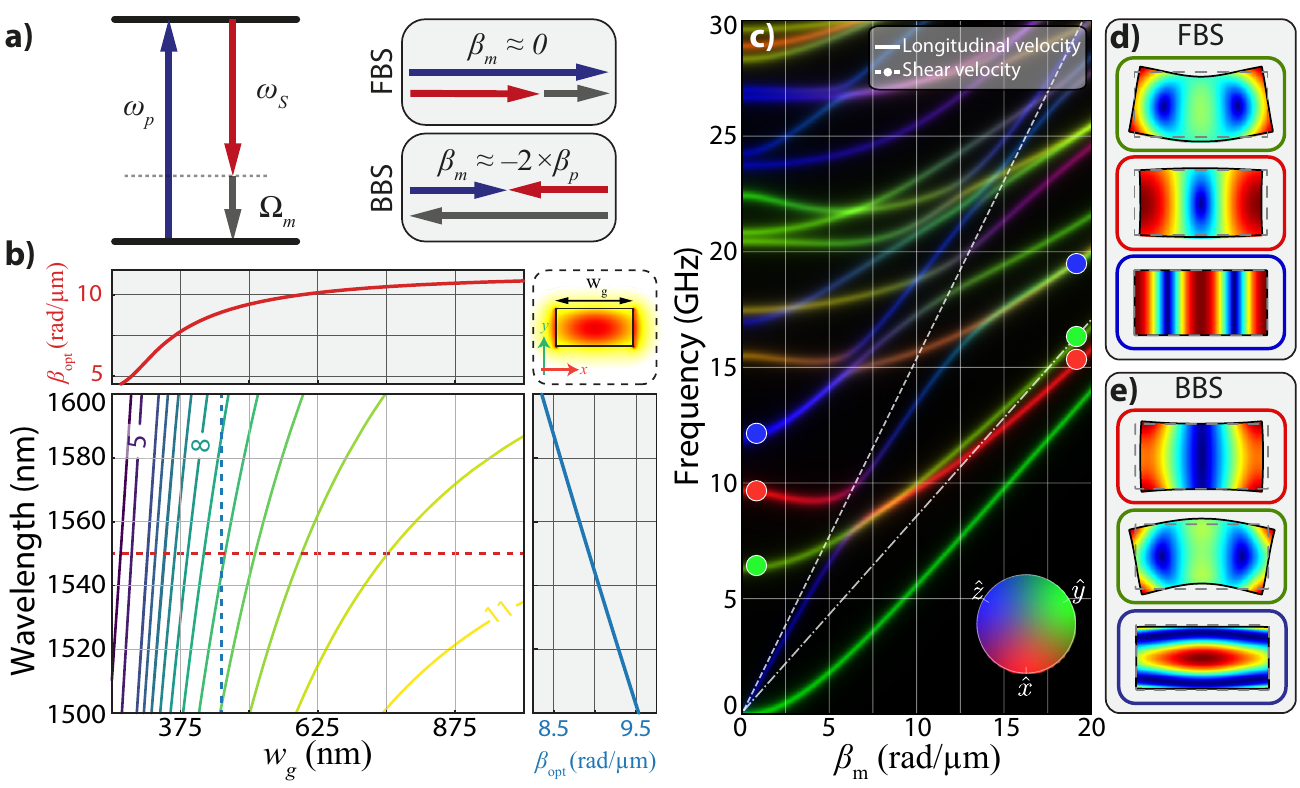}
    \caption{\textbf{Stimulate Brillouin Scattering}; \textbf{a)} an incident signal ($\omega_P,\beta_P$) is scatter by a mechanical wave ($\Omega_m,\beta_m$) to a scattered stokes mode ($\omega_S, \beta_S$), the frequency shift between the two optical modes is given by $\Omega_m$, with mechanical wave-vector satisfying the phase-matching condition ($\beta_m = \beta_P-\beta_S$). \textbf{b)} Optical wave-vector ($\beta_\mathrm{opt}$ in $\SI{}{\radian/\um}$) of a rectangular silicon waveguide as a function of the waveguide width and input wavelength. The top right inset shows the fundamental quasi transverse-electric mode (TE-like) considered for the plots. In all cases the waveguide thickness is kept in $\SI{220}{\nm}$. The red and blue dashed lines are cuts in the isocontour plot along the waveguide width and input wavelength, respectively. The top panel (solid red line) shows the optical wave-vector as a function of the waveguide width for a fix input wavelength of $\SI{1550}{\nm}$. The rate in which the wave-vector changes with the waveguide width for $w_g=\SI{450}{\nm}$ is (not show) $\partial \beta_\mathrm{opt}/\partial w_g\approx\SI{11.5e-3}{\radian/(\um.\nm)}$. The left panel (solid blue line) shows a linear dispersion for the optical wave-vector as a function of the input wavelength for $w_g=\SI{450}{\nm}$. \textbf{c)} Mechanical dispersion for the $y$-symmetric modes of a silicon rectangular waveguide ($\SI{450}{\nm}$ wide and $\SI{220}{\nm}$ tall), the displacement ratio along each Cartesian direction is given by a primary color ($x$: Red; $y$: Green; $z$: Blue) which are then combined to represent the polarization of each mode. \textbf{d)} and \textbf{e)} Mode profiles of the total displacement for representative modes highlighted in the dispersion curves in \textbf{c)} for $\beta_m = 0$, and $\beta_m = \SI{20}{\radian/\um}$ respectively.}
    \label{fig:disp}
\end{figure}

Forward and Backwards Brillouin Scattering are defined by the relative propagation between the incident and scattered optical modes, and are the two possible configurations that satisfy the phase matching condition in a waveguide geometry. For simplicity we will consider in this work the case of intramodal scattering, where the interacting optical modes comes from the same dispersion branch. 
Considering the phase-matching condition, illustrated in \figref{fig:disp}{(a)}, we can safely assume that the mechanical modes interacting via FBS are at the cut-off frequency regardless of the incident optical wave-vector, while in the case of BBS their mechanical wave-vector is roughly twice the value of the optical mode, meaning that the dispersion relations of mechanical modes can be no longer ignored.

In order to gain some insight of the optical and mechanical modes involved in the BBS and FBS, we use finite element method (FEM) to simulate the optical and mechanical dispersion of a floating rectangular waveguide as shown in \figref{fig:disp}{(b)-(c)}. For a typical waveguide width and thickness used in CMOS compatible devices (waveguide width: \SI{450}{nm}; silicon layer thickness: \SI{220}{\nm}), the  optical wave-vector for the fundamental TE-like (inset \figref{fig:disp}{(b)}) is $\beta_\mathrm{opt}\approx \SI{9}{\radian/\um}$  in the telecom C-band. Two other important features arise from \figref{fig:disp}{(b)}. First, we notice that above $w_g=\SI{450}{\nm}$ the optical wave-vector is mildly affected by the waveguide width, since the rate in which the wave-vector changes with the waveguide width ($\partial\beta_\text{opt}/\partial w_g$) has its maximum around $w_g=\SI{300}{\nm}$ (not show) and then monotonically decreases for larger waveguide width. This is an important feature when considering the geometrical dispersion in the BBS~\cite{wolff2016brillouin}. Second, by changing the pump frequency we can fine tune the optical wave-vector allowing for fine tuning of the BBS mechanical frequency.

\figref{fig:disp}{(c)} shows the mechanical frequency dispersion for a waveguide width $w_g=\SI{450}{\nm}$. Mechanical modes are colored according to the displacement ratio for each Cartesian direction, where red, green, and blue colors are related to $x$, $y$, and $z$-dominant polarized modes; the polarization dependence is clearly seen for the $\beta_m=0$ mechanical modes shown in \figref{fig:disp}{(d)}.

Other than the phase matching condition, the Brillouin optomechanical gain over the scattered mode ($G_B$) -- often called Stokes seed in pump-probe experiments -- strongly depends on the transverse profile of both optical and mechanical modes by the expression:
\begin{equation}
G_B (\Omega) = Q_m \frac{2\omega_p\mathcal{L}(\Omega)}{m_\text{eff}\Omega_m}\left| \int f_\text{PE}\,dA  + \int f_\text{MB} \, dV \right|^2 \,,
\label{eq:gain}
\end{equation}
\noindent where $\Omega$ is the beating frequency between the pump signal and the stokes seed, $Q_m$ is the mechanical quality factor, or Q-factor, $m_\text{eff}$ is the effective linear mass density of the mechanical mode, $\mathcal{L}$ is a Lorentzian function centered at $\Omega_m$ and with full-width-half maximum (FWHM) given by the mechanical linewidth ($\Gamma_m=\Omega_m/Q_m$), and $f_\text{PE}$ ($f_\text{MB}$) is the optomechanical overlap due to the photoelastic (moving boundaries) effect~\cite{Wiederhecker2019}. The coupling factor $G_B/Q_m$ can be calculated by numerically computing the mode profile of both optical and mechanical modes using COMSOL Multiphysics~\textregistered~ FEM simulations. In the case of silicon waveguides however, we can already expect that $x$-polarized ($z$-polarized) mechanical modes should have larger gain for FBS~\cite{santos2017hybrid} (BBS~\cite{Espinel2017}); these modes can be readily identified using \figref{fig:disp}{(c)}.

As we depart from the cut-off region in the mechanical dispersion ($\beta_m>q0$), anti-crossing and hybridization become more common adding another layer of complexity to our problem. A relevant example would be the $z$-polarized first order longitudinal mode (L$^{(1)}$)~\cite{Fraser1957,Krushynska2011}, highlighted in blue in \figref{fig:disp}{(e)}, which at $\beta \simeq \SI{14}{\radian/\um}$ hybridizes acquiring a Lamb-like character with significant displacement in the $y$-direction.

\section{Design}
\begin{figure}[ht!]
    \centering
    \includegraphics[width=1\columnwidth]{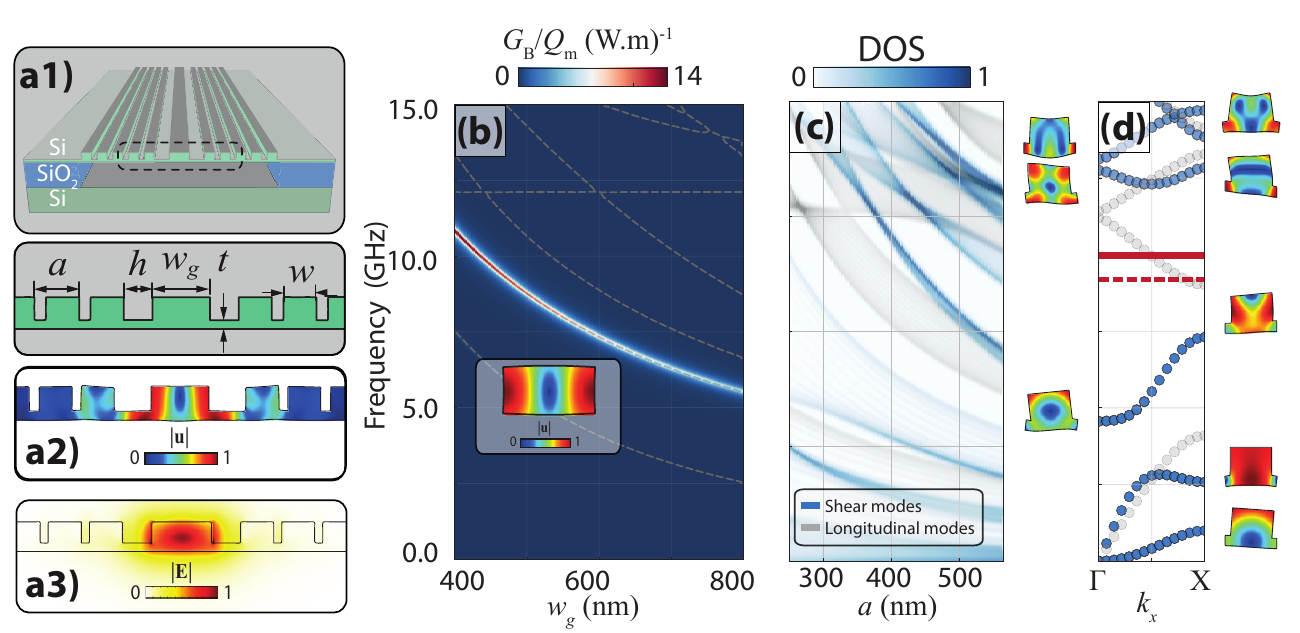}
    \caption{\textbf{FBS waveguide design process}. \textbf{a)} Schematics for the proposed waveguide. \textbf{(a1)} Cross section of the waveguide highlighting the main geometrical parameter. \textbf{(a2)} mechanical and \textbf{(a3)} optical modes profile for the mechanical breathing-like mode and fundamental TE-mode for $w_g =\SI{450}{\nm}$,  $h = \SI{220}{\nm}$, $a = \SI{310}{\nm}$, $w = \SI{250}{\nm}$. \textbf{b)} coupling factor ($G_B/Q_m$) colormap as a function of the waveguide width for modes with $\beta_m = 0$. $y$-symmetric mechanical modes are highlighted by the dashed lines to aid the visualization of modes with very low coupling factors. Inset: profile of the L$^{(2)}$ mechanical mode (breathing-like) that yields the highest coupling factor with the optical TE-mode. The linewidth of mechanical modes was arbitrary chosen based on aesthetics. \textbf{c)} Density of States (DOS) for the PhnC's shear (blue shades) and longitudinal (gray shades) modes with $\beta=0$ as a function of the lattice parameter ($a$) for $w/a = 0.8$. Lighter (darker) region corresponds to lower (higher) mode density. \textbf{d)} Band diagram for a 1D-PhnC ($a = \SI{340}{\nm}$, $w = \SI{250}{\nm}$), the shear modes are shown in blue and the longitudinal modes in grey dots. The red solid line represent the frequency of the L$^{(2)}$ mode from a silicon nanowire ($w_g = \SI{450}{\nm}$) and the dashed represent the frequency for a L$^{(2)}$-like mode of the waveguide shown in \textbf{(a2)}. Inset: mode profile for the crystal modes with non zero frequency at the points $\Gamma$ and $X$, only one unit cell is shown.}
    \label{fig:fwd}
\end{figure}

To mimic the previous floating waveguide, we propose a device that consists of a central rectangular waveguide on top of a suspended membrane that is mechanically isolated from the substrate by a 1-D phononic crystal cladding (PhnCC). \figref{fig:fwd}{(a)} shows a schematic of the design highlighting the main geometric parameters. 
We choose to fix both the central rib and the PhnCC at the same heights, so the whole mechanical structure can be fabricated in a single dry etching step. This is particularly important since multiple steps usually compromise the alignment between the nanostructures. We also opted to work around minimal support thickness since the Brillouin optomechanical gain would approach the one from the ideal floating silicon nanowire. So we considered for our simulations a Si thickness of $\SI{220}{\nm}$ and a etching depth of $\SI{160}{\nm}$  ($t=\SI{60}{\nm}$), as well as other dimensions compatible with commercial foundries rules, which already proven to produce ultra-high optical and mechanical quality factor devices~\cite{Benevides2017, santos2017hybrid}.

\subsection{FBS design}
We separated the design process for this structure between the Brillouin Active Region (BAR), and the Phononic Crystal Cladding (PhnCC). We approximate the BAR to a fully suspended rectangular waveguide; this is a valid approximation since in a ridge waveguide most of the Brillouin optomechanical gain occurs in the central portion, which has a higher optical field density. This approximation makes the BAR design much simpler -- as long as the waveguide support does not significantly affects the optical and mechanical modes, which is the case for $t=\SI{60}{\nm}$.

In Fig.~\ref{fig:fwd} we show the design process of a waveguide optimized to harness intramodal FBS using the fundamental TE-like mode. Similar discussion was thoroughly addressed for silicon bullseye-type cavity~\cite{santos2017hybrid} and silicon nitride waveguides~\cite{Dehghannasiri2017}. The Brillouin  optomechanical interaction tends to increase for tighter confinement of both optical and mechanical modes, as shown by the coupling factor ($G_B/Q_m$) map in \figref{fig:fwd}{(b)}, however, excessive compactness has been reported to deteriorate the optical propagation losses~\cite{Tran2018}. In order to maintain a lower optical loss we kept $w_g = \SI{450}{\nm}$ in our design.

For the considered geometrical parameters mode that yields the highest intramodal FBS $G_B/Q_m$, is the L$^{(2)}$-Mode (breathing-like)~\cite{Fraser1957,Krushynska2011} around \SI{10}{\GHz}. To confine this mode is necessary to tailor the phononic crystal's parameters to create a bandgap that encompasses this frequency. Since this mode is planar at cut-off($u_x\neq0$, $u_y\neq0$ and $u_z=0$)~\cite{Krushynska2011}, a partial bandgap considering only modes that are also planar is sufficient to confine it, this makes the waveguide design within our fabrication constrains relatively simple.
In \figref{fig:fwd}{(c)} we show the evolution for the density of states (DOS) of the PhnC as a function of the lattice parameter ($a$), where the filling factor ($w/a$) is kept constant. It is clear from this graph the relation between the mechanical bands frequency and $a$, showing that we can easily find a bandgap of several GHz around the desired frequency just by tuning $a$. 
\begin{figure}[ht!]
    \centering
    \includegraphics[width=1\columnwidth]{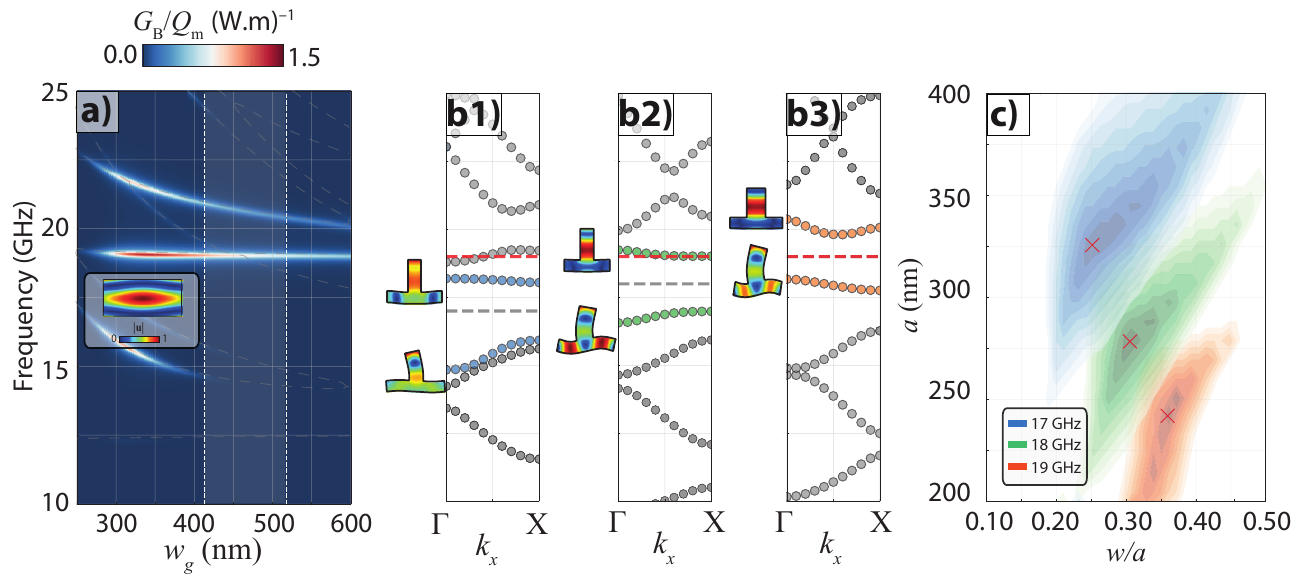}
    \caption{\textbf{BBS waveguide design process}. \textbf{a)} coupling factor ($G_B/Q_m$) map as a function of of the waveguides width at $\beta_m = \SI{18}{\radian/\um}$, every mode is highlighted by the dashed lines to aid the visualization of modes with low coupling factors. The region where the optical frequency lies inside the photonic C-Band is represented by the shading. The linewidth of the mechanical modes was arbitrary chosen based on aesthetics. Inset: profile of the mode, L$^{(1)}$ (Lamb-like), that yields the highest coupling factor in this region. \textbf{b)} Band diagram for PhnCs marked on the map shown in \textbf{(c)} at $\beta_m = \SI{18}{\radian/\um}$. The bandgap under study is highlighted by its color and its central frequency is represented by the gray dashed lines. The red dashed lines represents the frequency of the lamb-like mode at this particular value of $\beta_m$. Inset: mode profile for the relevant crystal modes at the $\Gamma$ point, only one unit cell is shown. \textbf{c)} The evolution of the bandgaps highlighted in \textbf{(b)} is shown in the PhnC parameters map, the shading represents the relative depth of the bandgap considering a particular central frequency, each level is spaced by 1/10 of the maximum depth at this central frequency. The 'x' marking represents the PhnCs whose band diagrams are shown in \textbf{(b)}.}
    \label{fig:bwd1}
\end{figure}

As the PhnCC is attached to the BAR, both the mechanical and optical modes of the BAR might hybridize with modes from the support slab. In the case of the optical mode the slot effect can be avoided by making the support width ($h$) larger than the mode penetration depth in the air region, which is roughly $\SI{120}{\nm}$ in our case. Due to the large $x$-displacement at the edges, the $L^{(2)}$-mode easily couples to the slab's dilatational mode, like $L^{(2)}$ this mode can be understood as an acoustic Fabry-Perot mode and so the mechanical frequency decreases with the slab width ($h$ in this case).For certain values of $h$ we expect to see crossing between the $x$-polarized dilatational and the $y$-polarized shear modes. Regions near these crossing points should be avoided in order to preserve the mechanical mode integrity. 

\figref{fig:fwd}{(d)} shows the band-diagram for the final phononic crystal design together with the FBS frequency for the BAR structure (red solid lines) and the resulting FBS frequency for the waveguide structure depicted in \figref{fig:fwd}{(a2)} (red dashed lines). The frequency difference between the BAR and final structure is small when compared to the bandgap width, not affecting the final gain values significantly, thus supporting our design strategy for FBS.

\begin{figure}[ht!]
    \centering  
    \includegraphics[width=1\columnwidth]{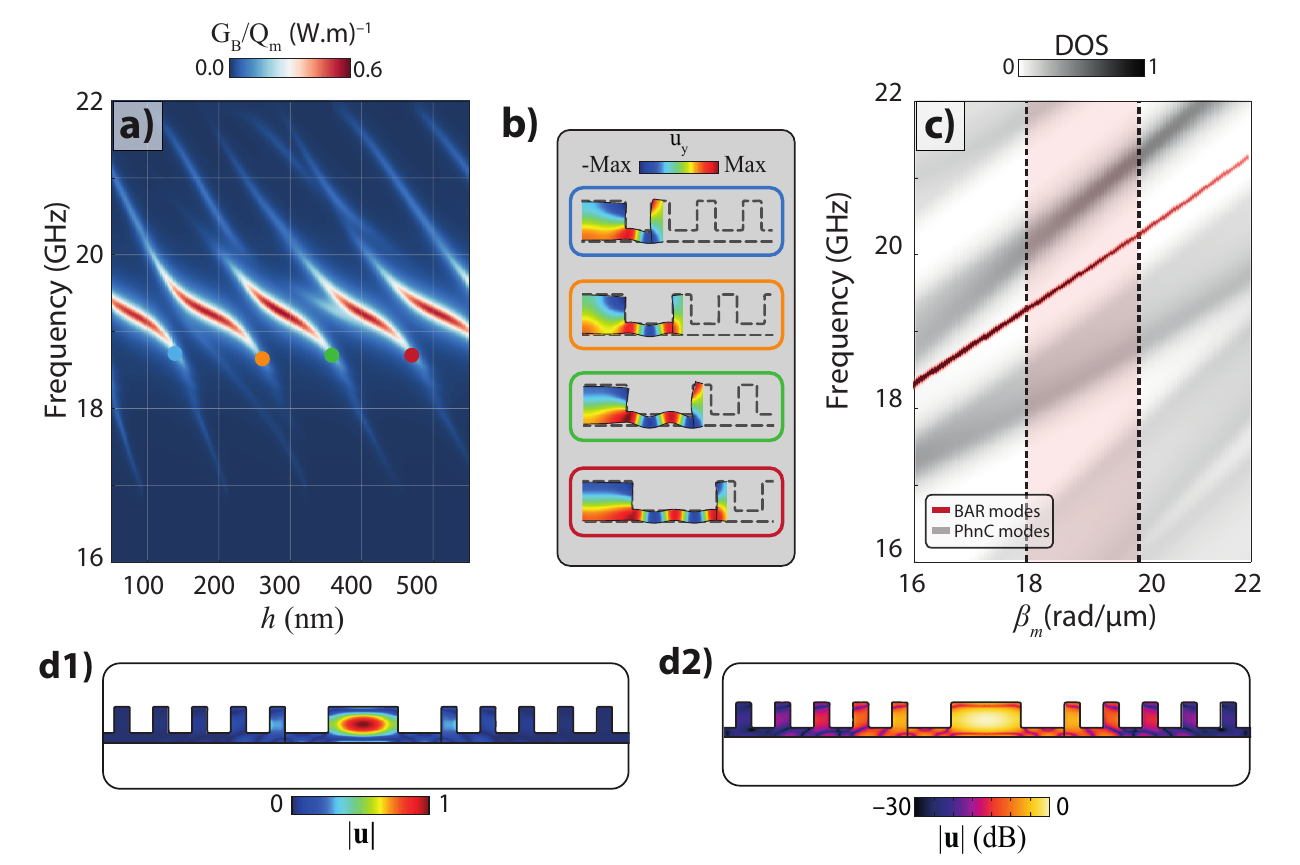}
    \caption{\textbf{BBS waveguide process}. \textbf{a)} coupling factor ($G_B/Q_m$) colormap for the complete structure ($w_g = \SI{450}{\nm}$,$a = \SI{250}{\nm}$, $w= \SI{100}{\nm}$) as a function of the support slab width ($h$) considering $\beta_m = \SI{18}{\radian/\um}$. \textbf{b)} Mechanical displacement along the $y$-direction for the highlighted modes at (top to bottom) ($h = \SI{140}{\nm}$, $h = \SI{260}{\nm}$, $h = \SI{360}{\nm}$, $h = \SI{460}{\nm}$). \textbf{c)} The mechanical dispersion for the first order lamb mode of the complete structure ($w_g = \SI{450}{\nm}$,$h = \SI{280}{\nm}$, $a = \SI{250}{\nm}$ and $w= \SI{100}{\nm}$), the shaded area shows the mechanical wave-vector tuning range provided by changes in the optical frequency inside the photonic C-band. Mode profile of the first order lamb mode of the complete structure in the linear scale \textbf{d1)} and in the dB scale \textbf{d2)}. The linewidth of mechanical modes in \textbf{(a)} and \textbf{(c)} was arbitrary chosen for aesthetic purpose.}
    \label{fig:bkw2}
\end{figure}

\subsection{BBS design}
For BBS the design process has to be modified to carefully consider non zero $\beta_m$. As discussed before the mechanical wave-vector is now coupled to the geometry via the optical dispersion \figref{fig:disp}{}. To ease the bandgap search in this case, we fix the wave-vector for the mechanical simulations; this could be experimentally achieved by varying the optical frequency. In \figref{fig:bwd1}{(a)} we evaluate the Brillouin optomechanical gain of a floating rectangular waveguide for intramodal BBS as a function of $w_g$ considering $\beta_m =\SI{18}{\radian/\um}$, the shaded region in the map is the width range where the optical frequency lies inside the photonic C-band and therefore can be easily adjusted.

BBS favors mechanical modes that have significant displacement in the $z$-direction~\cite{Espinel2017}, and so the mode that yields the highest coupling factor is the $yz$-polarized L$^{(1)}$-mode (Lamb-like), as seen in the inset of \figref{fig:bkw2}{(a)}. As discussed before, this is a hybrid mode with significant displacement in the $z$ and $y$ direction, and so a full bandgap is necessary to confine it to the BAR region. Full bandgaps are expected to be narrower, demanding for a finer design control over the bandgap central frequency and width, as seen in \figref{fig:bwd1}{(b)}. To obtain this level of control we consider one extra degree of freedom for the optimization: the filling factor ($w/a$). \figref{fig:bwd1}{(c)} shows the evolution of the bandgaps highlighted in \figref{fig:bwd1}{(b)} considering both geometric parameters: $a$ and $w/a$. Darker colors refers to larger bandgaps at the respective central frequency. By following the evolution of the bandgap position and size as a function of the central frequency we can easily choose the most appropriate parameters for the best confinement of the BAR mode. 

The Lamb-like mode is susceptible to crossings with the shear-vertical modes from the support due its small displacement in $y$-direction. Considering the silicon bulk sound velocity we expect the support shear modes to have a wavelength of $\sim\SI{300}{\nm}$ around $\sim\SI{19}{\GHz}$. As a results, for every half wavelength increase in the support width ($h$) we expect to see a crossing of these two modes. This is precisely the origin for the anti-crossings seen in \figref{fig:bkw2}{(a)}, where we calculate the Brillouin optomechanical gain for the full waveguide structure at $\beta_m =\SI{18}{\radian/\um}$ as a function of the width $h$. The mechanical mode displacement profiles with low gain are shown in \figref{fig:bkw2}{(b)} with $h-$size difference roughly equal to half the shear mode wavelength. The small discrepancy between the expected $h-$size difference and values recovered from the FEM simulations, occurs since the slab defined by $h$ does not have fixed boundary conditions at both ends, effectively increasing its width.

Now that the geometric parameters are set we can turn back our attention to the mechanical dispersion curves. In \figref{fig:bkw2}{(c)} we have both the dispersion diagram of the PhnCC (shaded gray) and BAR modes (shaded red). The slope of the Lamb-like mode is close to shear modes, resulting in a small relative frequency shift between the defect and crystal modes. Therefore a mode well centered inside the bandgap for a given wave-vector, will then stay centered throughout a large $\beta_m$ range, this provides a tuning range for the mechanical frequency of \SI{1}{\GHz} for optical frequencies within the photonic C-band. 

\section{Discussion}

Acoustic damping arises from three main sources: geometry dependent clamping losses, fabrication dependent resonance broadening~\cite{Safavi-Naeini2019,wolff2016brillouin} and intrinsic mechanical losses~\cite{Zener1939,Chavez-Angel2014}. The latter is dominated by thermal effects such as Thermo-Elastic Damping (TED) and the Akhiezer Effect (AKE). Those are out of the scope of our design, but can be avoided in low-temperature experiments.

Mechanical clamping losses arises due to the lower acoustic velocity of the substrate material in comparison to the device material, in our design this effect is successfully suppressed by blinding the mechanical mode from the device-substrate interface. In both Breathing-like mode and Lamb-like mode, respectively shown in \figref{fig:fwd}{} and \figref{fig:bkw2}{}, the mechanical displacement after only 5 PhnC cells was attenuated over 30 dB, this leads to radiation limited Q-factors of $Q_\text{rad}>10^5$ for the breathing-like and $Q_\text{rad}>10^6$ for the Lamb-like. Is important to note that those values are expected to increase exponentially with the addition of more cells~\cite{santos2017hybrid}.

The broadening of the Brillouin resonance due to fabrications defects have two length scales: the first being short-range variations, where broadening arises from surface roughness and composition, and the second being long-range  where slow variations  on the cross-section geometry lowers the mechanical coherence~\cite{wolff2016brillouin}. The first one should not limit the quality factor considering the ultra-high mechanical Q ($>10^{10}$) already reported in silicon nanocavities at low temperatures~\cite{MacCabe2019}. In the second case its effect can be estimated by calculating the sensitivity of the mechanical frequency towards geometric variations. As expected the breathing-like mode is more sensitive to width variations ($S_{w_g}\cong\SI{9}{\MHz/\nm}$) while the Lamb-like mode is more prone to height variations ($S_{t}\cong\SI{29}{\MHz/\nm}$). A table containing the sensitivity of the each mode towards different geometric parameters is presented in the Appendix A. Considering that the record width (height) variations on \SI{300}{\mm} diameter wafer processing are on the order of \SI{0.3}{\nm} (\SI{0.15}{\nm})~\cite{Selvaraja2014,van2015interaction} then the breathing(Lamb)-like the mechanical quality factor is limited at $Q_m\cong1725$ $(\cong2200)$ which would lead to a Brillouin Gain of $G_B\cong\SI{7750}{\per\watt\per\meter}$ $(\cong\SI{1200}{\per\watt\per\meter})$. The calculation of the fabrication limited $Q_m$ is discussed in appendix B.

An interesting comparison can be make between our design and the ideal fully suspended silicon nanowire with the same BAR dimensions. Tab.~\ref{tab:gain_values} presents both the coupling factor ($G_B/Q_m$) and the dominant geometric sensitivity ($S_\text{max}$) for the two waveguides:

\begin{table}[ht!]
    \centering
    \begin{tabular}{c|c|c|c}
        Interaction & Waveguide &  $G_B/Q_m  (\SI{}{\per\watt\per\meter})$ & $S_\text{max}(\SI{}{\MHz/\nm})$ \\
        \hline
         \multirow{2}{*}{FBS} & Fully Suspended & 10.91& 21\\
         & PhnC Assisted & 4.50& 9\\
         \hline
          \multirow{2}{*}{BBS} & Fully Suspended & 0.83 & 48\\
           & PhnC Assisted & 0.54& 29
    \end{tabular}
    \caption{Impact of the PhnCC on forward and backward Brillouin gain.}
    \label{tab:gain_values}
\end{table}


Even with the addition of the PhnCC the effective mechanical domain in our design is still larger than in the ideal case due to the connecting slab and the penetration depth of the mechanical mode inside the crystal. From the table above we can see that this increase manifest itself in two ways: a decrease in the Brillouin optomechanical coupling factor but also a decrease in the mechanical frequency sensitivity to changes in the geometry. If we consider that acoustic losses are limited by fabrication then we have $Q_m\propto S_\text{max}^{-1}$ (see Appendix B), meaning that the decrease in the coupling factor is balanced by the increase in the mechanical coherence, resulting in a  Brillouin optomechanical Gain ($G_B$) that is effectively the same as the ideal case (\SI{96}{\%} for FBS, \SI{110}{\%} for BBS). Our design then provides a feasible and robust structure, that can support itself over large lengths without halting the interaction, while still having the same gain performance of a fully suspended nanowires.

\section{Conclusion}

The presented waveguide design promotes an alternative approach for realizing SBS in silicon integrated waveguides. The addition of the mechanical shields (PhnCC) successfully mitigates mechanical clamping losses homogeneously throughout the propagation axis while still preserving the ease of fabrication. For both FBS and BBS optimized designs our waveguide possesses Brillouin Gain equivalent to fully suspended silicon nanowires with an additional advantage of long unperturbed interaction lengths making it ideal for interactions that possesses long acoustic decay lengths such as intramodal BBS. This approach could open the possibilities for the implementation of BBS based devices and high-frequency experiments in integrated silicon photonics. Our phonon shield design in principle could also enable the addition of electrical contacts without affecting too much on the mechanical mode. This could open novel applications such as thermal tuning, and p-i-n type structures that could lower the free-carrier lifetime and two-photon absorption in the waveguides~\cite{Gajda2011}. Although we focus on silicon substrate, others material that suffer from mechanical clamping losses such as silicon nitride and gallium arsenide should also benefit from the presented design. 

\section*{Data Availability}
FEM and scripts files for generating each figure are available at the ZENODO repository (10.5281/zenodo.4148337)~\cite{zenodo_brillouin_waveguide}. 

\section*{Funding}
This work was supported by S\~{a}o Paulo Research Foundation (FAPESP) through grants 2019/13564-6, 2018/15580-6, 2018/15577-5, 2018/25339-4, Coordena{\c c}\~ao de Aperfei{\c c}oamento de Pessoal de N{\'i}vel Superior - Brasil (CAPES) (Finance Code 001), Conselho Nacional de Desenvolvimento Cient\'{i}fico e Tecnol\'{o}gico through grants 425338/2018-5, 310224/2018-7, and Financiadora de Estudos e Projetos (Finep).

\section*{Appendix A: Sensitivity to dimensional variations}

Inhomegenities over the waveguide cross section is one of the limiting factor for the Brillouin Resonance Quality factor, in order to calculate this limit is necessary to calculate each mode sensitivity to the dimensional variations. The sensitivities were calculated using COMSOL Multiphysics~\textregistered~ FEM simulations, we swept each parameter over a \SI{5}{\nm} range and then linear fitting the mechanical frequency. In these simulations the optical wavelength was kept at \SI{1550}{\nm}. The table below show the sensitivities of the mechanical modes studied with respect to the dimensional parameters that should be more affected during the fabrication processes.

\begin{table}[ht!]
\centering
\begin{tabular}{|c|c|c|}
\hline
Parameter                       & Mode           & Sensitivty (MHz/nm) \\ \hline
\multirow{2}{*}{$W_g$}          & Breathing-like & 8.70                \\ \cline{2-3} 
                                & Lamb-like      & 5.74                \\ \hline
\multirow{2}{*}{$h$}            & Breathing-like & 3.86                \\ \cline{2-3} 
                                & Lamb-like      & 5.94                \\ \hline
\multirow{2}{*}{$t$} & Breathing-like & 0.81                \\ \cline{2-3} 
                                & Lamb-like      & 28.92               \\ \hline
\end{tabular}
\end{table}

\section*{Appendix B: Structural Broadening Calculations}

The Brillouin gain over the stokes seed in a single-pass waveguide is given by the equation bellow:

\begin{equation}
    (P_\text{out}/P_\text{in})_{dB} = \frac{10}{\log(10)}\int_0^LP_\text{Pump}(z)\, G_B(\Omega,z) \, dz
\end{equation}

As a result of the inhomogenities over the waveguide cross-section the frequency of the phase-matched mechanical mode is now a function of the longitudinal parameter $z$. Considering that each parameter is perturbed over a mean value, the frequency can be written as:

\begin{equation}
\label{eq:omega_m}
    \Omega_m(z) = \Omega_0 +\sum_i \xi_i(z)S_i\,,
\end{equation}

\noindent where $\Omega_0$ is the unperturbed frequency, $\xi_i$ is the perturbation over the $i$ dimensional parameter and $S_i$ is the frequency sensitivity related to the $i$ parameter.
In a realistic approximation the dimensional variations are randomly distributed along the entire waveguide length with values within the tolerance of the fabrication procedure ($\xi_{i,\text{max}}$). If the characteristic length of those variations is smaller than the effective length of the Brillouin interaction ($L_\text{eff}$) and the pump power stays relatively constant throughout the length of the waveguide the above equation can be simplified, and the optical frequency offset ($\Omega$) dependency then reads:

\begin{equation}
\label{eq:nlorentzian}
\int_{-\infty} ^{\infty}\frac{\Gamma^2/4}{(\Omega-\Omega_m)^2+\Gamma^2/4} P(\Omega_m) \, d\Omega_m \,,
\end{equation}

\noindent where $P(\Omega_m)$ is the probability of finding $\Omega_m$ as that the frequency of the phase-matched mechanical mode at any given value of $z$. The probability density function (PDF) of $\Omega_m$ can be obtained by doing the convolution of the PDFs of every term in equation. Due to the nature of the convolution operation, \equaref{eq:nlorentzian} is dominated by the larger of the parameters $\Delta\Omega_m = \xi S$. For a limit of vanishing intrinsic linewidth ($\Delta\Omega_m \gg \Gamma$) the resulting linewidth tend to $2\Delta\Omega_{m,\text{max}}$ ($Q_m \rightarrow \Omega_m/(2\Delta\Omega)$).


\end{document}